%% file: main.tex
\documentclass[sigconf,natbib=false,nonacm]{acmart}

\AtBeginDocument{%
  \providecommand\BibTeX{{%
    \normalfont B\kern-0.5em{\scshape i\kern-0.25em b}\kern-0.8em\TeX}}}

\usepackage{import}
\usepackage{graphics}
\graphicspath{{imgs/}}

\import{sections/}{00-preamble}

\import{numbers/}{numbers-generated}
\import{numbers/}{numbers}

\usepackage{appendix}
\usepackage{booktabs}
\usepackage{multirow}

\begin{document}
\date{}

% \title{How Do You Choose Your AI Component? An Interview Study of Secure AI Integration in Practice}
\title{From Adoption to Deployment: A Qualitative Study on AI Integration in Software Development Practice}

\author{Mahzabin Tamanna}
  \affiliation{%
  \institution{{North Carolina State University}
  \country{USA}}}
\email{mtamann@ncsu.edu}

\author{Elizabeth Lin}
  \affiliation{%
  \institution{{North Carolina State University}
  \country{USA}}}
\email{etlin@ncsu.edu}
\author{Sparsha Gowda}
  \affiliation{%
  \institution{{North Carolina State University}
  \country{USA}}}
\email{ssgowda3@ncsu.edu}

\author{Laurie Williams}
  \affiliation{%
  \institution{{North Carolina State University}
  \country{USA}}}
\email{lawilli3@ncsu.edu}

\author{Dominik Wermke}
  \affiliation{%
  \institution{{North Carolina State University}
  \country{USA}}}
\email{dwermke@ncsu.edu}

\begin{abstract}
\import{sections/}{01-abstract}

\end{abstract}


\thispagestyle{plain}
\pagestyle{plain}

\maketitle
\section{Introduction}\label{sec:intro}
\import{sections/}{02-intro}

\section{Background and Related Work}\label{sec:related-work}
\import{sections/}{03-related-work}

\section{Research Methodology}\label{sec:method}
\import{sections/}{04-method}

\section{Results}\label{sec:results}
\import{sections/}{05-results}

\section{Discussion}\label{sec:discuss}
\import{sections/}{06-discussion}

\section{Conclusions}\label{sec:conclusion}
\import{sections/}{07-conclusion}

\begin{acks}
This work was supported and funded by the National Science Foundation Grant No. 2207008. Any opinions expressed
in this material are those of the author(s) and do not necessarily
reflect the views of the National Science Foundation. We thank our participants for their participation in the study, for sharing their experiences, and for providing valuable input towards securing AI components as part of the software supply chain. 
\end{acks}
\printbibliography
\end{document}

%% file: sections/00-preamble.tex
\usepackage[utf8]{inputenc} % should not be necessary with current LaTeX versions
\usepackage[T1]{fontenc}

\usepackage[dvipsnames,table]{xcolor}

\usepackage[normalem]{ulem}
\usepackage{fontawesome}
\usepackage{csquotes}
\usepackage{setspace}
\usepackage{lineno}
% Figures and graphics
\usepackage{svg}
\usepackage{graphicx}
\usepackage{graphics}
\usepackage{caption}
\usepackage{subcaption}
\usepackage{float}
\usepackage{tcolorbox}
\usepackage{enumitem}
\usepackage{ellipsis}

\usepackage{tikz}
\usetikzlibrary{shapes,arrows,fit,matrix,positioning,decorations.pathreplacing,arrows.meta}

% Fonts
\usepackage{times}
\usepackage[gen]{eurosym}

% Hyperref MUST come before hyperxmp
\usepackage{hyperref}
\usepackage{hyperxmp}

% Tables
\usepackage{tabularx}
\usepackage{makecell}
\usepackage{booktabs}
\usepackage[para, flushleft]{threeparttable}
\usepackage{array}
\usepackage{adjustbox}
\usepackage{multirow}
\usepackage{tablefootnote}

\expandafter\let\csname c@tblerows\endcsname\rownum

\usepackage{enumitem}

\usepackage{etoolbox}

\newcolumntype{R}[2]{%
    >{\adjustbox{angle=#1,lap=\width-(#2)}\bgroup}%
    l%
    <{\egroup}%
}
% no 
\usepackage{xurl}
\PassOptionsToPackage{hyphens,spaces,obeyspaces}{xurl}

\usepackage[
	backend=biber,
	isbn=false,
	%doi=false,
	% sorting=none,
	sortcites=true, % sort wihtin \cite commands
	%url=false,
	%eprint=false,
	%defernumbers=true,
	%mincitenames=1,
	%maxcitenames=1,
	maxbibnames=99, % ensure a full list of authors in the bibliography, prevents printing only the first x 
	%uniquelist=false,
	%datamodel=acmdatamodel,
	%style=acmnumeric,
	%citestyle=numeric-comp
	]{biblatex}
\addbibresource{ref.bib}

\DeclareCiteCommand{\citeauthorfull}
    {\boolfalse{citetracker}%
    \boolfalse{pagetracker}%
    \DeclareNameAlias{labelname}{given-family}%
    \usebibmacro{prenote}}
    {\ifciteindex
        {\indexnames{labelname}}
        {}%
        \printnames{labelname}}
        {\multicitedelim}
{\usebibmacro{postnote}}

\usepackage{orcidlink}

% commands to use across paper

\newcommand{\myparagraph}[1]{\vspace{0.25em}\noindent\textbf{#1.}}

% \newcommand{\omit}[1]{\ldots}
%command to hide in tex file and show dots in pdf
\newcommand{\hide}[1]{\ldots}

% Quote formatting
\AtBeginEnvironment{quote}{\vspace{-0.5em}}
\AtEndEnvironment{quote}{\vspace{-0.5em}}

\newcommand{\participantQuote}[2]{{\color{Orchid}\textit{``#1''}~({#2})}}

\usepackage{xparse}
\usepackage{xstring}
\newcommand{\rangecap}[1]{%
  \IfValueTF{#1}{%
    \IfEqCase{#1}{%
      {1}{\textit{A few}}%
      {2}{\textit{A few}}%
      {3}{\textit{A few}}%
      {4}{\textit{Some}}%
      {5}{\textit{Some}}%
      {6}{\textit{Some}}%
      {7}{\textit{Many}}%
      {8}{\textit{Many}}%
      {9}{\textit{Many}}%
      {10}{\textit{About half}}%
      {11}{\textit{About half}}%
      {12}{\textit{About half}}%
      {13}{\textit{The majority of}}%
      {14}{\textit{The majority of}}%
      {15}{\textit{The majority of}}%
      {16}{\textit{Most}}%
      {17}{\textit{Most}}%
      {18}{\textit{Most}}%
      {19}{\textit{Almost all}}%
      {20}{\textit{Almost all}}%
      {21}{\textit{Almost all}}%
      {22}{\textit{All}}%
    }[\textbf{Unknown Value}] % Default case
  }{\textit{No Value Given}}% If no value provided
  \ignorespaces % Suppress space after the command
}

% usage example: \range{12}, \rangecap{13}

\newcommand{\range}[1]{%
  \IfValueTF{#1}{%
    \IfEqCase{#1}{%
      {1}{\textit{a few}}%
      {2}{\textit{a few}}%
      {3}{\textit{a few}}%
      {4}{\textit{some}}%
      {5}{\textit{some}}%
      {6}{\textit{some}}%
      {7}{\textit{many}}%
      {8}{\textit{many}}%
      {9}{\textit{many}}%
      {10}{\textit{about half}}%
      {11}{\textit{about half}}%
      {12}{\textit{about half}}%
      {13}{\textit{the majority of}}%
      {14}{\textit{the majority of}}%
      {15}{\textit{the majority of}}%
      {16}{\textit{most}}%
      {17}{\textit{most}}%
      {18}{\textit{most}}%
      {19}{\textit{almost all}}%
      {20}{\textit{almost all}}%
      {21}{\textit{almost all}}%
      {22}{\textit{all}}%
    }[\textbf{Unknown Value}] % Default case
  }{\textit{No Value Given}}% If no value provided
  \ignorespaces % Suppress space after the command
}

\usepackage{fancyhdr}
% Configure the header style
\fancypagestyle{workingmanuscript}{
    \fancyhf{} % Clear all header and footer fields
    \fancyhead[C]{\textcolor{red}{\sffamily \Huge Working Manuscript - Not Intended for Distribution}} 
     % Remove the horizontal line under header
}

%% file: numbers/numbers-generated.tex
% DO NOT include numbers manually here, this file will be overwritten
% (Add them to 'numbers.tex' instead)

%% file: numbers/numbers.tex
% Manual numbers file
% You can include your numbers manually here

% VAR format
% Definition: \definevar{name}{XX}
% Usage in document: \var{name} prints "XX"
% e.g., \definevar{participants}{321}

%% file: sections/01-abstract.tex
 The increasing adoption of Large Language Models (LLMs) as Artificial Intelligence (AI) components in modern software systems introduces distinct security risks to the software supply chain. 
While many considerations and safety mechanisms are in place for components of the traditional software supply chain, the recent rapid adoption of AI components and platforms has overlooked these hard-learned lessons.
Selecting and integrating AI models without clear guidance on how these choices affect system security may leave applications vulnerable to threats, such as malicious components, data leakage, and unintended behavior.~\textit{The goal of this study is to understand practitioners' decision-making process and security considerations in selecting and integrating AI components through an exploratory semi-structured interview study.}
Toward this goal, we conducted semi-structured interviews with 22 software developers, architects, and AI practitioners across diverse organizations about how they integrate AI components into their software.

Our analysis finds that practitioners’ model selection is predominantly driven by functional criteria, including performance, accuracy, cost, and specific features, e.g., tool-calling or multimodal support, while security is rarely considered as an evaluation criterion. We observe a consistent lack of security concern throughout the AI component integration process, with established software supply chain lessons overlooked or ignored. The industry is repeating the historically costly mistakes of early software dependency management, prioritizing rapid reuse and availability over security and provenance. We distill our findings into actionable recommendations for AI adopters, model providers, and researchers, advocating for a proactive, security-by-design approach that integrates security evaluation into component selection and sustains it throughout the software development lifecycle.

%% file: sections/02-intro.tex
Following recent advances in artificial intelligence (AI), AI components such as large language models (LLMs) and other machine learning models are increasingly integrated into software systems.
This widespread integration introduces security and privacy risks through the models themselves and expands the systems' software supply chains, including external models and services.

According to a 2026 Gartner report, \$2.5 trillion was spent globally for integrating AI into software~\cite{gartner2.5T}, with 72\% of enterprise software now including at least one AI-powered feature, driving a 73\% increase in AI API usage~\cite{servicenow,yahoofinance}.
The 2025 Stack Overflow survey reports 84\% of respondents are using or plan to use AI tools in their software development, representing a 76\% increase from the previous year~\cite{stack2025}.
Industry analysts projected the trajectory will continue, predicting that at least 55\% of software engineering teams will actively build LLM-based features into products by 2027~\cite{gartner}.

This integration, however, introduces additional attack surfaces.
Embedding third-party AI models, orchestration layers, and agent frameworks into software systems creates supply chain dependencies~\cite{wang2025large,kezron2024securing,williams2022discovery}.
A 2026 cyber report found that 73\% of 1,500 security practitioners experienced AI-related threats that significantly impacted their organizations~\cite{darktrace}, while IBM reported 13\% of 600 global organizations have suffered an AI-related data breach~\cite{IBMCost}.
Moreover, these risks are compounded by insecure developer practices.
For example, the National Cybersecurity Alliance (NCA) reports that 43\% of developers have exposed sensitive information to AI systems, including financial and client data~\cite{NCA}.
Recent incidents, such as the LiteLLM~\cite{litellm} and OpenClaw~\cite{openclaw} compromises, demonstrate that vulnerabilities in upstream AI packages can result in credential theft and broader downstream system compromises.
Recognizing the threat, governing bodies such as the National Institute of Standards and Technology (NIST)~\cite{nist} and the National Cyber Security Center (NCSC)~\cite{ncsc} issued regulatory guidance framing third-party models as high-risk supply chain dependencies.

With AI models being integrated as critical components of software systems, the industry is effectively facing a~\textit{Back to the Future} scenario in which the same ignorance has arisen again. Based on our findings, the current industry approach to AI adoption and integration mirrors the historically costly handling of software dependencies. For decades, practitioners have prioritized rapid reuse and readily available components while overlooking components as critical elements of software supply chain security and potential attack surfaces~\cite{larios2020selecting,lysenko2025select,zhan2021research,tanzil2024people}. 
Subsequent empirical work demonstrates that unvetted, reusable components introduce significant attack vectors and systemic risk, contributing to widespread vulnerabilities across ecosystems, including new attack classes, such as dependency confusion~\cite{neupane2023beyond,spracklen2025we,zheng2024towards,alfadel2023empirical,gu2024survey}.

% Prior work context
Previous work in the context of AI security has largely focused on benchmarking model effectiveness~\cite{mohammadi2025evaluation,lee2025sec}, secure code generation~\cite{tihanyi2025secure,li2024fine,kavian2024llm,nazzal2024promsec,lin2025give}, or evaluating models' efficacy for specific security tasks such as adversarial testing~\cite{he2023large,shen2025pentestagent}, fuzzing~\cite{jiang2024fuzzing,zhou2024liftfuzz}, vulnerability analysis~\cite{sheng2025llms,yang2025code}, threat modeling~\cite{elsharef2024facilitating,jedrzejewski2025thremolia}.
We argue that there remains a critical gap in the literature regarding how industry practitioners actually select, evaluate, and manage third-party AI models within their development workflows.

\textit{The goal of this study is to understand practitioners’ decision-making process and security considerations in
selecting and integrating AI components through an exploratory
semi-structured interview study.}
% Bridging the gap
Towards this goal, we conducted an exploratory semi-structured interview study of software practitioners.
We engaged 22 industry professionals, including software engineers, team leads, and AI engineers, to investigate their experiences, security postures, and decision-making processes when integrating AI components into software systems.  Our interview study was guided by the following research questions:

\begin{enumerate}[topsep=1pt, leftmargin=*, itemsep=2pt, label=\textbf{RQ\arabic*}:, ref=\textbf{RQ\arabic*}]
\item \textit{What factors influence practitioners' decisions to select AI models as a software component?}
We want to identify how practitioners select an AI model to integrate into software development:
(i) What technical, organizational, and security-related aspects are prioritized?
(ii) How do practitioners assess and operationalize these influencing factors during the LLM selection process?
(iii) What is the model selection decision process?

\item \textit{What challenges do practitioners face when integrating AI models as a software component?} 
We want to identify challenges and experience-based insights about the barriers, risks, and constraints encountered during AI integration in software systems.

\item \textit{How security risks are perceived and what safeguards are applied to mitigate the security risks?} 
We want to understand practitioners' interpretations, awareness, and prioritization of risks for integrating AI models into systems. We also want to identify the security controls and safeguards deployed to reduce risk in industry projects.
\end{enumerate}

Guided by these research questions, through a qualitative interview study, we aim to provide a grounded understanding of current AI component and model adoption practices and identify areas for improvement to support the secure adoption, development, and deployment of AI models as parts of the software supply chain.

\myparagraph{Highlights}
We observe that security remains consistently under-emphasized and less prioritized throughout the AI component integration process, often secondary to considerations of functionality, availability, and ease of reuse:
(i) The selection of AI models is predominantly driven by functional requirements, cost, use-cases, and ecosystem constraints, with security rarely considered as an evaluation or selection metric, (ii) Because intrinsic model vulnerabilities are largely ignored, practitioners default to reactive, infrastructure-level safeguards such as input/output filtering and isolated environments, and (iii) The opaque, black-box nature of AI models, along with constant version updates complicate long-term system stability and cause technical debt.
The prevailing functionality-first approach to AI model adoption risks repeating the same software supply chain failures observed with traditional dependencies.
Avoiding this outcome requires a shift toward proactive, secure-by-design practices that integrate security evaluation into component selection and sustain throughout the software development lifecycle.

The remaining paper is structured as follows: Section~\ref{sec:related-work} provides background and related work, and Section~\ref{sec:method} explains the methodology and outlines the interview structure. Section~\ref{sec:results} presents the results of the interview study, Section~\ref{sec:discuss} provides a detailed discussion and recommendations. Lastly, Section~\ref{sec:conclusion} outlines the conclusions.

%% file: sections/03-related-work.tex
In this section, we discuss prior work most relevant to our interview study on the adoption and integration of AI components, focusing on selecting software components, as well as selection and threats of AI models as components.

\myparagraph{Selecting Software Components}
The selection and management of software components has been the subject of prior research~\cite{geisterfer2006software, fahmi2009study, gholamshahi2019software, carvallo2007determining, grau2004descots}. A growing body of research has studied the consideration of security in selecting software components~\cite{wermke2022committed, wermke2023always, zahan2023software, pashchenko2018vulnerable, pashchenko2020qualitative}.
Nevertheless, several challenges persist in this domain.
Such challenges include, but are not limited to, the security of open-source libraries~\cite{zimmermann2019small, alfadel2023empirical, wang2020empirical}, incompatibilities arising from dependency version updates~\cite{lin2025context}, and the abandonment of maintained dependencies~\cite{miller2025understanding}. Prior work has examined open-source package ecosystems, including PyPI~\cite{guo2023empirical, bommarito2019empirical, alfadel2023empirical}, npm~\cite{decan2018impact, sejfia2022practical}, and Maven~\cite{mir2023effect, balliu2023challenges,tamanna2025your, wang2020empirical}.
With the growing integration of AI into software systems, AI models can increasingly be regarded as software components in their own right, warranting the same scrutiny as traditional dependencies, such as open-source libraries. Our work follows this reasoning, yet, to our knowledge, no prior study has examined how practitioners actually select, evaluate, and manage AI models as components within this framing, nor what challenges arise in the process.

\myparagraph{AI Components}
Software developers can use and integrate AI in various ways, and there are also multiple approaches to integrating AI into software systems.
One common use is through AI assistants embedded within code editors, which allow developers to interact directly with an LLM without leaving their development environment.
Prior work has explored how users interact with AI assistants~\cite{ross2023programmer} and compared the impact of AI assistants on code quality~\cite{perry2023users}.
A related but distinct use of AI is the AI-powered code editor, such as Cursor~\cite{cursor} and GitHub Copilot~\cite{copilot}, which provides context-aware code generation by understanding the broader codebase. 
The use of these AI-powered editors has motivated studies that aim to understand their use and challenges~\cite{sergeyuk2025using,chaudhary2026large, mcnutt2023design}.
Beyond developers using AI assistants and code editors, AI can also be integrated through direct API calls to LLM providers. 
Given that each provider exposes different interfaces, third-party services, such as OpenRouter~\cite{openrouter}, have emerged to provide universal APIs that abstract over individual providers.
A more sophisticated form of AI integration is the use of agents, which leverage LLM reasoning to perceive environments, plan, and execute multi-step actions with little or no human oversight.
 While prior work has cataloged these integration modalities, it has largely treated them in isolation and has not examined how practitioners navigate the practical decision-making involved in selecting and integrating AI components across these forms. Our study addresses this by capturing practitioner experience across integration.

\myparagraph{Selection and Threats of AI Components}
% With the rapid advancement of large language models (LLMs), their integration into software systems has become increasingly prevalent.
A growing body of research has explored the adoption of AI in software development~\cite{negri2024systematic, li2024user, denny2024computing, tabarsi2025llms, hu2024always, khan2024most}.
Mink et al.\cite{mink2023everybody} examined the benefits and pain points of ML-related security tools among security practitioners.
With the growing adoption of AI assistants, \citeauthor{klemmer2024using}~\cite{klemmer2024using} found that although their use is prevalent, users tend to mistrust AI assistants and use them with care.
\citeauthor{tabarsi2025llms}~\cite{tabarsi2025llms} looked into the larger-scale impact LLMs have on software development processes.
In addition to exploring how AI tools shape software development, prior work has also addressed the guidance for adoption.
\citeauthor{lee2024don}~\cite{lee2024don} focused on AI privacy and found a lack of guidance around AI privacy. More broadly, \citeauthor{khati2025mapping}~\cite{khati2025mapping} mapped practitioner trust in LLMs across software engineering tasks, surfacing tensions between perceived utility and reliability. Collectively, these studies focus primarily on AI as a productivity tool rather than as an integrated system component subject to architectural, dependency, and security considerations. The security implications of AI integration have also received growing attention.
Several survey papers have examined the security challenges with LLMs and AI agents~\cite{deng2025ai, das2025security,tamanna2025security, yi2024jailbreak, dong2024attacks}. Prompt injection is among the most recognized and readily executable attacks targeting AI systems and has been the subject of numerous prior studies~\cite{greshake2023not, zhan2024injecagent, liu2023prompt, chen2025struq}. 
The threat landscape is particularly evident in the case of AI agents, where the consequences of such attacks are amplified by their autonomous decision-making. A number of studies have examined attacks and defenses associated with AI agents~\cite{zhang2024agent, chen2024agentpoison, andriushchenko2024agentharm, yu2025survey}, highlighting the expanded attack surface and risks introduced by integrating autonomous AI components.
The expanded attack surface motivates our work to explore and understand the use of AI components in software systems and identify areas for improvement. Despite these advances in work, a clear gap remains. Prior studies have examined either the usability of AI tools or the technical vulnerabilities of AI systems, but have not investigated how practitioners experience the end-to-end integration of AI components into their software development. Our study addresses this gap, offering grounded, practitioner-level insight into a problem space that has received substantial technical attention but limited empirical, human-centered investigation.

%% file: sections/04-method.tex
In this section, we provide the methodology of our study.
To answer our RQs, we conducted iterative, semi-structured interviews, followed by qualitative coding between October 2025 -- March 2026.
We employed a qualitative research design to examine how practitioners select and integrate AI components, focusing on decision-making processes, barriers, and security considerations that are not directly observable.
Semi-structured interviews provide depth and flexibility, enabling in-depth exploration while accommodating emerging themes, which is particularly suitable given the evolving nature of AI integration.
We next describe participant recruitment, interview design and procedure, coding and analysis, as well as ethical considerations and study limitations.

\subsection{Interview Structure}~\label{interview_structure}
We developed a structured interview guide following our research questions, focusing on three core areas:
(1) AI component selection criteria,
(2) challenges with AI integration, and
(3) perceived security risks and safeguards.
The interview guide was iteratively refined through feedback from the research group and four pilot interviews with both researchers and industry practitioners. These interviews were used exclusively to refine the interview guide and associated protocols. As such, the data and participants are not included in the final dataset of 22 interviews.

Based on the initial pilot interviews, we made minor adjustments, including clarifying question wording, adding follow-up questions to improve coverage of model integration and security-related practices, and establishing a more natural conversational flow during the interview process. 
\input{Tables_and_Figures/interview_structure}.The interview guide was organized into five main sections, each comprising opening questions and targeted follow-ups for deeper insights.
We present the final interview structure in Figure~\ref{fig:interview_flow} and also discuss in detail below.
We began each of the 22 interviews by providing a brief overview of our research objectives, the purpose of the interview, and the general structure. We also collected signed electronic forms and verbal consent from the participants. 

\myparagraph{1. Projects and Context}
The first section aimed to obtain each participant's background and project context.
Participants were asked to describe their role, the systems they work on, and how AI fits into their development workflows. This section served three purposes: (1) to ease participants into the interview by starting with familiar and descriptive topics; (2) to obtain contextual grounding for subsequent questions, and (3) to obtain participants' demographic information. 

\myparagraph{2. AI Model Selection and Consideration}
In the second section, we examined how practitioners choose among AI models.
We asked participants about their model selection criteria, including technical, organizational, and security-related factors.
We further explored model evaluation practices, including the model assessment process, the tools employed, and the checkpoints.
We also covered how practitioners operationalize evaluation and compliance in practice, including compliance requirements and company policies.
This section aims to identify the factors that practitioners consider when selecting AI components to integrate into their software systems.

\myparagraph{3. AI Model Integration Challenges}
In the third section, we asked participants about challenges or barriers encountered during AI integration to capture real-world complexities, including technical, organizational, and security issues.
We also explored incidents or failures, including incorrect model outputs, unexpected behavior, and system-level breakdowns. Additionally, we asked participants about AI model-related inconveniences encountered in their projects.
When discussing potentially sensitive topics, such as past incidents, we encouraged participants to speak in general terms as needed.

\myparagraph{4. Security Risk and Safeguard Considerations}
In our fourth set of questions, we covered how practitioners perceive and mitigate the risks of integrating an AI component.
We asked participants about perceived security risks (e.g., hallucinations, prompt injection, data leakage), the safeguards they have applied to secure the development pipeline, their monitoring practices, and access controls throughout the development pipeline. We also asked about guidance and documentation for including or excluding models in their system.

\myparagraph{5. Improvement and Reflection}
In the final section, we asked participants to reflect on current practices and identify gaps in existing tools, frameworks, or processes
We explored the desired evaluation metrics and benchmarks, the missing tooling for safe and reliable integration, organizational improvements, and their reflections for the future of AI components.
This section provided insights into how the ecosystem could evolve to better support practitioners.

\myparagraph{Outro} 
Following the interview sections, we asked participants to share any additional insights or aspects that were not covered or that they wished to elaborate on. Finally, we thanked them for their time and contributions, offered an opportunity to ask final questions or provide comments, and concluded with a brief debriefing.

\subsection{Participant Recruitment}
We recruited participants for our interview study using purposive sampling to identify practitioners with relevant experience in adopting and integrating AI components.

\myparagraph{Recruitment Channels} We recruited participants for our study through several channels to cover a diverse set of industry experiences and backgrounds. We leveraged Upwork\footnote{https://www.upwork.com/}, a professional hiring platform, to recruit practitioners with hands-on experience in LLM-based or AI-driven projects. We enhanced our sample by reaching out to experienced professionals through posting on social media and community working groups on platforms such as Slack and Discord.
To further expand our participant pool, we also employed snowball sampling, asking interviewees to recommend other relevant practitioners in their professional networks. This method enabled us to reach qualified, experienced participants, including those who may not be easily accessible through other channels.

\myparagraph{Inclusion-Exclusion Criteria}
Participants were recruited using purposive sampling, selecting those with relevant experience with AI components, such as large language models, in software development. For the study, we required participants to (1) be currently employed or recently engaged in software development, machine learning, AI engineering, or related roles, (2) have direct, hands-on experience in integrating AI within development, and (3) be sufficiently fluent in English for an interview. Participants with experience across different stages of the software development lifecycle were prioritized to capture diverse perspectives.
Individuals were excluded if they lacked industry experience with AI, were not involved in technical roles, or were unable to provide informed consent. Additionally, participants from regions where compensation or data collection posed regulatory or logistical constraints were excluded to maintain compliance with ethical and institutional requirements.

To verify these criteria, we employed a two-stage screening process. First, a sign-up form in which participants were required to provide examples of AI-integrated projects, specifying the models used and their roles in the projects. We used these descriptions to measure 'relevant experience,' specifically looking for evidence of system integration, model evaluation, or architectural decision-making. This allowed us to distinguish qualifying hands-on experience from 'theoretical knowledge or casual use,' such as using a public chatbot for non-development tasks. Candidates with qualifying experience are then invited to an interview, where we further confirm their background and project context during the initial phase of the discussion. Following this screening process, 22 participants were recruited for the study. Each represented a distinct project and organization and was compensated with \$40 in Amazon.com vouchers or the Upwork equivalent for their time.

\subsection{Interview Procedure and Data Collection}
Prior to participation in the study, all individuals were provided with an informed consent form outlining the study’s purpose, study procedures, data handling procedures, compensation, IRB information, and their rights as participants. Electronic consent was formally obtained from each participant before proceeding with the interview.
We emphasized that participation was entirely voluntary, participants could decline to answer any question, they could withdraw at any time, and the study focused on understanding practices rather than evaluating or judging their (security) decisions.
Furthermore, we explicitly assured participants that all data would be anonymized.
After addressing any preliminary questions, we began the recording and conducted introductory questions about the participant’s role, experience, and project descriptions. Interviews were conducted remotely as video calls via Zoom and lasted approximately 45–60 minutes (individual times listed in Table~\ref{tab:demogarphics}). In instances where the discussion exceeded 60 minutes, participants were informed that continuing was entirely voluntary and would not affect their compensation.

\subsection{Coding and Analysis}
% Transcription
After conducting the interviews, all audio recordings were transcribed using a locally deployed~\textit{OpenAI Whisper} model to maintain data privacy. The first author manually reviewed and corrected all transcripts to address transcription inaccuracies. To analyze the interview data, we employed an iterative thematic analysis approach~\cite{terry17}.

\myparagraph{Codebook}
We developed an initial codebook using a hybrid approach following the interview structure in Section~\ref{interview_structure}, as well as identifying potential codes from pilot interviews. The codebook's sections correspond to the sections described in our interview guide.

\myparagraph{Coders \& Coding Workflow}

The coders are three researchers with prior experience in qualitative analysis and knowledge of AI as components and software development practices. From the pilot phase, the first author conducted exploratory coding on the initial four interviews and added new subcodes to refine the codebook. For the main coding phase, the three authors coded the transcripts iteratively as they came in, with the second and third authors dividing the odd and even transcripts. We ensured each transcript was independently coded by at least two coders, a main coder and one additional coder from the research team. Every time a new code was introduced, the three coders discussed the addition and modified the codebook as needed. The codebook was iteratively refined throughout the analysis to incorporate emerging codes for themes and insights from subsequent interviews.
All three coders contributed to this iterative refinement process through discussion to achieve a shared understanding of the coding scheme and codebook.

\myparagraph{Conflict Resolution}
Following the independent coding of each transcript, the coders met to compare their assigned codes, discuss discrepancies, and reach a consensus. The first author led the conflict resolution, directly addressing minor differences (e.g., missing subcodes or slight variations in marked words) and updating the cookbook based upon discussions and agreements. This iterative process continued until all codes had been applied to all transcripts and no unresolved disagreements remained. To maintain consistency, previously coded transcripts were revisited
by the coders and selectively re-coded where necessary, using the updated codebook. As all discrepancies were reconciled through this process, we do not report inter-rater reliability (IRR), maintaining consistency with prior work using similar approaches~\cite{mcdonald2019reliability,lin2025context,sb2025they}.

\myparagraph{Stopping Criteria}
We observed that the number of subcodes increased rapidly during the initial interviews and stabilized in later interviews.
We concluded the interviews when we reached thematic saturation~\cite{francis2010adequate,guest2020simple}, defined as the point at which no new codes or substantive insights emerged across the last four consecutive interviews.
We report the codes per interview in Table~\ref{tab:demogarphics}.

\myparagraph{Thematic Analysis} Once the coding was finalized, we engaged in a process of thematic mapping. We grouped related codes into broader sub-themes and overarching themes that captured the practitioners’ perceptions and practices. This involved reviewing the coded excerpts to ensure the themes accurately represented the data and the research questions. We discuss identified themes in Section~\ref{sec:results}.
\subsection{Ethical Consideration:}
We structured our interview study to follow the ethical
principles outlined in the Menlo report. Prior to conducting interviews, we obtained approval for our study setup from our university’s Institutional Review Board (IRB). The primary consideration is the
protection of participants identity and professional context. We mitigate
potential risks by anonymizing all transcript data, removing identifiable details from quotes, and reporting bucketed demographics. All data was collected, handled, and stored in an institutional secure storage.

\subsection{Limitations}
As with most qualitative interview studies, our findings are based on self-reported practices and may be subject to recall bias or subjective interpretation.
We mitigated this threat by following up with questions and identifying patterns across a diverse participant pool rather than relying on individual perspectives. The process of qualitative coding introduces subjectivity. We mitigated this threat through multi-coder analysis, iterative codebook refinement, and consensus-based conflict resolution.

While our participants span diverse roles, experience levels, and domains, the use of purposive sampling may introduce selection bias and limit representativeness.To mitigate sampling limitations, we recruited participants across multiple channels, applied explicit inclusion criteria, and complemented our approach with snowball sampling to reach practitioners who were less accessible through conventional recruitment methods.
Although we followed a systematic recruitment and analysis process, additional interviews may have revealed new insights.
We mitigated this threat by continuing data collection until thematic saturation was reached, defined as the point at which no new codes or substantial insights emerged across the last four consecutive interviews.
Overall, while inherent limitations of qualitative research remain, we adopted established best practices in interview design, data collection, and throughout our analysis to ensure the consistency and credibility of our findings and the study.

%% file: Tables_and_Figures/interview_structure.tex
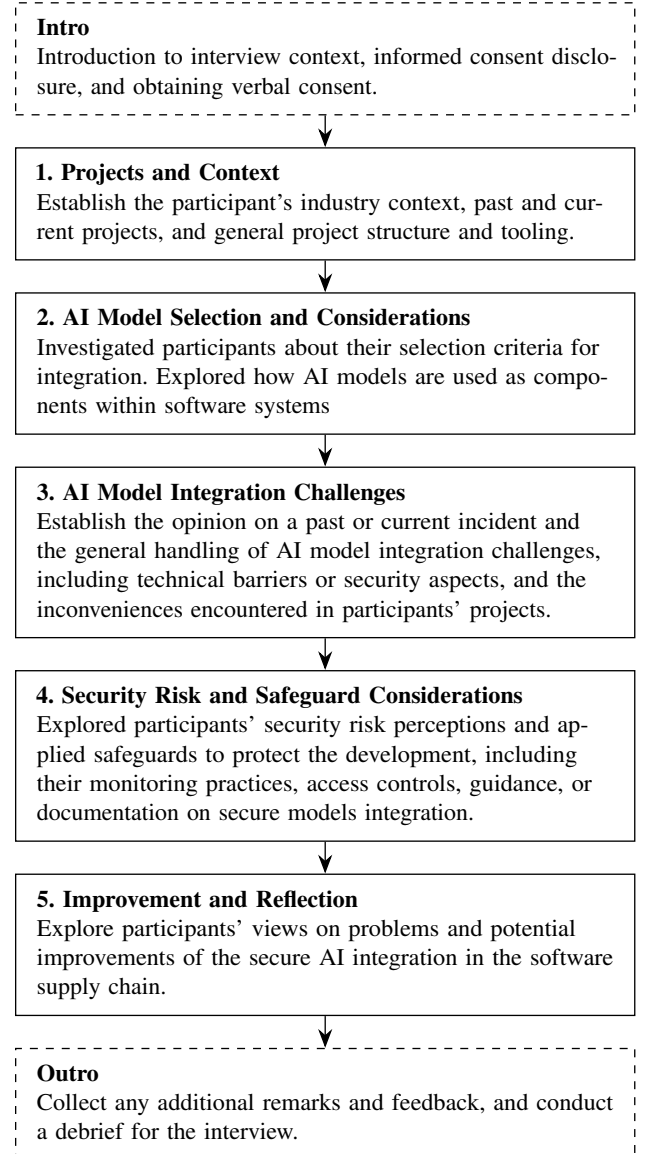
\begin{figure}[tbp]
\centering
\begin{tikzpicture}[
    font=\normalsize,
    >={Stealth[length=2.5mm]},
    node distance=0.42cm,
    solidbox/.style={
        draw=black,
        rectangle,
        line width=0.5pt,
        align=left,
        inner xsep=8pt,
        inner ysep=6pt,
        text width=0.90\columnwidth
    },
    dashedbox/.style={
        draw=black,
        dashed,
        rectangle,
        line width=0.5pt,
        align=left,
        inner xsep=8pt,
        inner ysep=6pt,
        text width=0.90\columnwidth
    },
    flow/.style={
        -{Stealth[length=2.5mm]},
        line width=0.5pt
    }
]

\node[dashedbox] (intro) {%
\textbf{Intro}\\
Introduction to interview context, informed consent disclosure, and obtaining verbal consent.
};

\node[solidbox, below=of intro] (s1) {%
\textbf{1. Projects and Context}\\
Establish the participant's industry context, past and current projects, and general project structure and tooling.
};

\node[solidbox, below=of s1] (s2) {%
\textbf{2. AI Model Selection and Considerations}\\
Investigated participants about their selection
criteria for integration. Explored how AI models are used as components within software systems
};

% \node[solidbox, below=of s2] (s3) {%
% \textbf{3. AI model Integration}\\
% Identified participants about their selection
% criteria, including performance, cost, latency, privacy, and ease
% of integration.
% };

\node[solidbox, below=of s2] (s3) {%
\textbf{3. AI Model Integration Challenges} \\
Establish the opinion on a past or current incident and the general handling of AI model integration challenges, including technical barriers or security aspects, and the inconveniences encountered in participants' projects.
};

\node[solidbox, below=of s3] (s4) {%
\textbf{4. Security Risk and Safeguard Considerations}\\
Explored participants' security risk perceptions and applied safeguards to protect the development, including their monitoring
practices, access controls, guidance, or documentation on secure models integration.
};

\node[solidbox, below=of s4] (s5) {%
\textbf{5. Improvement and Reflection}\\
Explore participants' views on problems and potential improvements of the secure AI integration in the software supply chain.
};

\node[dashedbox, below=of s5] (outro) {%
\textbf{Outro}\\
Collect any additional remarks and feedback, and conduct a debrief for the interview.
};

\draw[flow] (intro.south) -- (s1.north);
\draw[flow] (s1.south) -- (s2.north);
\draw[flow] (s2.south) -- (s3.north);
\draw[flow] (s3.south) -- (s4.north);
\draw[flow] (s4.south) -- (s5.north);
% \draw[flow] (s5.south) -- (s6.north);
\draw[flow] (s5.south) -- (outro.north);

\end{tikzpicture}
\caption{Overview of the interviews' flow and topics. After introducing each section with a general question, we followed-up with specific questions (if not already covered). Due to our semi-structured interview approach, participants were allowed to diverge from this flow at any time.}
\label{fig:interview_flow}
\end{figure}

%% file: sections/05-results.tex
In this section, we present the findings from our 22 semi-interviews with software developers, architects, and AI practitioners regarding the integration of AI components.
The reporting structure largely follows the interview guide outlined in Section~\ref{interview_structure}, with key insights summarized after each question block.
For quantitative context and reading flow, we indicate the number of participants in each section using the ranges defined in Figure~\ref{fig:interview_ranges}, e.g., ``\range{14}'' corresponds to 13--15 participants.
Participant quotations are reported verbatim, with de-identification, minor grammatical corrections for readability, and omissions denoted using brackets (\enquote{\textelp{}}).

\input{Tables_and_Figures/Scale_diagram}

\subsection{Participants and Projects}\label{Subsec:Participants}
In total, we recruited a multidisciplinary cohort of 22 practitioners for semi-structured interviews.
As detailed in Table~\ref{tab:demogarphics}, the participants span a wide range of roles, industrial sectors, and years of experience.
Interviewed participants included software developers and engineers (8), AI/ML engineers (6), data scientists (3), and individuals in strategic leadership or architectural roles (5), such as CEOs, AI architects, and project managers.
Experience levels within the group are diverse, from early-career practitioners with one to three years of tenure to veteran industry professionals possessing over 10 years of experience. \rangecap{10}(10) of the participants reported 4 to 6 years of professional experience, and \range{4} (4) participants had over 10 years of experience (on average 20 years).
These practitioners have experience across operating agile startups and sole-developer contexts to large-scale enterprises in highly regulated sectors, including healthcare (4), finance (7), cybersecurity (6), and manufacturing (3). 
Participants described a diverse range of projects involving AI components, including chatbots, advanced applications such as AI-automated penetration testing, and the development of governance layers for AI safety, overall indicating a versatile integration of AI models within the software supply chain.

AI use among participants shows a multi-model approach, with several dominant proprietary and open-source models being integrated.
The most prevalent commercial models were OpenAI's GPT models (including different versions), mentioned by 12 participants; Anthropic's Claude, mentioned by 11; Google's Gemini, mentioned by 10; and Microsoft's Copilot, mentioned by 2.
On the more open model side, Meta's Llama series had been adopted by 6 participants, alongside mentions of models such as Qwen (4) and Deepseek (2).
Practitioners generally mention obtaining these models through four primary channels:
direct APIs, cloud providers, Hugging Face, and local serving tools like Ollama and vLLM to run models on private GPU infrastructure.
According to our participants, models are also leveraged for specialized tasks, including Retrieval-Augmented Generation (RAG), agent orchestration, intricate data extraction, and conversational AI, across diverse operational contexts. 

\input{Tables_and_Figures/Participant_table}
\begin{tcolorbox}[colback=blue!5!white]
\textbf{Summary (Participants and Projects):} Our participants represent a diverse and multidisciplinary cohort spanning a wide range of roles, experience levels, and industry sectors.
Participants reported working on multiple AI-driven projects across varying organizational contexts, from startups and individual development settings to large, highly regulated enterprises.
\end{tcolorbox}

\subsection{Selection of AI Component (RQ1)}
In this section, we present the findings for \textit{RQ1: What factors influence practitioners’ decisions to select AI
models as a software component?}
Our questions targeted three key areas in the component selection process:
(1) the prioritization of technical, organizational, and security factors;
(2) the assessment and evaluation processes; 
(3) the key selection actors in projects and organizations.
 
\subsubsection{Selecting AI Components} 
In this section, we discuss the factors and themes that influence practitioner selection decisions.  

\myparagraph{Technical Capabilities and Performance} In our interviews,~\range{22}(22) practitioners mentioned functional capacity and performance as major factors for selecting models, including accuracy, speed, latency, and the ability to produce concise outputs.
Participants also mentioned that they sometimes compromise on speed for better accuracy in infrastructure building, while others prioritize faster, cheaper models.
Additionally, models' logical reasoning and tool-calling are mentioned as necessary specialized tasks, e.g., P17 mentioned: \participantQuote{We check sometimes their evaluation results, how good they are at math \textelp{}.}{P17},
as well as models being good at performing better on distinct data: \participantQuote{We used models like Qwen and Llama because there were two things we wanted to support. One was simple logical question and answers \textelp{} and the other was how good in deep research for financial documents.}{P11}.

Model selection is also influenced by several practical considerations, including alignment with the intended use case, popularity (e.g., download metrics), and the capacity of the context window.
Practitioners also evaluate thinking processes and the ability to correct mistakes: \participantQuote{\textelp{} \textins*{T}he model has to have better thinking ability \textelp{} Whatever is good at thinking, we need it.}{P01}.
Participants also mention favoring models that are easy to integrate with existing programming languages and frameworks, noting that open-source options are often preferred because they integrate well with the environment they use: \participantQuote{We usually prefer open source because we can integrate different types of languages that we use.}{P12}.
The ability to digest non-English languages was also mentioned as relevant: \participantQuote{\textins*{W}e check the language capacity. Since I live in Turkey, and all documents are in Turkish language, we make sure that it works well in the Turkish language.}{P17}.
For reasons to reject a model, participants mention verbosity: \participantQuote{The more verbose a model is, the less I want to use it.}{P05}.
Additionally, the model being too defensive is also mentioned as a rejection criterion.

\myparagraph{Organizational Directives and Strategic Alignment} For~\range{9}~(9) participants, selection of models is driven by organizational preferences rather than pure technical merit.

E.g., P9 prefers an in-house solution: \participantQuote{The reason why I selected Llama is that we can have some self-hosting.}{P9}. 
Our participants also mentioned that the vendor licensing influences their decision. Sometimes, there is a client-side licensing policy for models, and practitioners need to follow that: \participantQuote{The company prefers we use in-house tooling whenever possible, instead of open source, because that comes with licensing constraints.}{P14}. Related, another participant noted: \participantQuote{We used Copilot for that because we have Copilot’s license. Otherwise, we could have gone with other models.}{P15}. 
Strategic partnerships and clients' demands also play a critical role, sometimes overriding performance considerations: \participantQuote{It depends on the project, the company, and the audience, all the people involved in the project and their partnerships.} {P05}.
Compliance with company-specific security and data-handling policies, such as HIPAA or GDPR, which are often required in fintech and healthcare, is mentioned as a factor when selecting an AI component: \participantQuote{GDPR requires the data to stay in the region. In those cases, we cannot use the Deepseek because data goes to the China.}{P19}. Data residency requirements could be decisive as participant mentioned~\participantQuote{ '[Gemini] was the only option that met the data residency requirement. So we finally chose Gemini.}{P19}, illustrating how regulatory constraints can narrow model selection to a single viable option.
Based on our interviews, larger, more mature organizations appear to employ dedicated teams more often to manage internal governance and directives. In addition, we observed that practitioners in highly regulated sectors appear more inclined to implement and enforce formal security policies.

\myparagraph{Security, Trust, and Policy Considerations}
For \range{20} (20) participants, trust in a provider is a primary driver.
Rather than conducting deep security analysis of the models, practitioners appear to often rely more on the reputation of the model vendors, peer recommendations, and the endorsement from senior engineers: \participantQuote{\textins*{I}t gave me some sort of reliability that like the companies which are leading in AI have created this model.}{P15}. This sentiment was widely echoed across our interview process, with practitioners noting that: \participantQuote{I can tell you, in the market a lot of people trust and following big guys, Google, Microsoft Copilot, everybody's behind AWS.}{P07}. 
However, this reliance on trust is not unconditional.
Geopolitical concerns are mentioned as strong deterrents.
Several practitioners explicitly avoided certain foreign models due to perceived national security risks, government restrictions, or general skepticism about the model's origins: \participantQuote{we don't use deep-seek models. Still, it's open source, but people don't believe it, because it's, made from China.}{P15}.
Participants also mention leaking information or having a history of data breaches as rejection criteria: \participantQuote{\textins*{I}f a particular model is known for leaking information, you don't want to choose that.}{P11}. 
Based on our interviews, practitioners exhibited high trust in established vendors and stated they would deprecate any model if the provider permitted training on proprietary data. However,\range{5} (5) participants regarded such data usage as a routine condition of service.\participantQuote{They [model providers] use data because they want  to train their models and make them better... nowadays, data is everywhere, if it's not sensitive.}{P06}

\myparagraph{Economic and Resource Constraints}
For \range{18}~(18) practitioners, the model selection process is also based on cost-effectiveness and infrastructural needs.
Participants highlighted a constant dilemma between deploying the most capable model and remaining within budget: \participantQuote{\textins{The model is} easy to use because the cost is somehow friendly.}{P8}.
High token and API costs were mentioned as a rejection criterion, especially when a cheaper or older model can achieve similar performance for a given task.
% Beyond pricing, hardware limitations also influence decisions:
% Resource availability appears to influence whether participants select managed cloud APIs or self-hosted models.
AI components that are too large to run on available infrastructure or that require specialized compute due to model weight were rejected in favor of more lightweight alternatives.
The computing power required to host local models was also mentioned as a financial barrier.
Consequently, participants mention that they rely on and have to outsource data to managed cloud services.

\subsubsection{Assessment and Evaluation Process}
In this section, we discuss the assessment and evaluation process used by participants to align components with their requirements.

\myparagraph{What is Evaluated}
\rangecap{7}~(7) participants mentioned that their evaluation baseline was established with a focus on accuracy and factuality of the model's output, assessing the frequency of hallucinations and the overall precision of generated content.

In parallel to functional correctness, \range{10} (10) participants mention evaluations related to adversarial and privacy risks.
Participants mentioned evaluation for risks like data leakage and the accidental exposure of private information (PII), including a small number considering prompt injection.
They also mentioned reviewing the model’s memory function to ensure it does not improperly share sensitive data from previous user sessions.
Emphasizing the importance of guardrails, P04 noted:
\participantQuote{We ensure that the model is actually aligned, so that you cannot provide any misleading data.}{P04}.
\rangecap{6}~(6) participants mentioned that deployment viability is further dictated by measuring technical efficiency through inference latency and hardware overhead (e.g., GPU and RAM utilization).

\rangecap{4}~(4) practitioners mentioned that for complex, agent-driven workflows, functional reliability is important.

In contexts such as customer support, \range{2}~(2) participants mention preferring human evaluation, prioritizing contextual relevance, faithfulness, and a consistent, empathetic, professional tone.

\myparagraph{Process of Evaluation} The process of evaluating models involves diverse methods, ranging from simple manual checks to structured automated pipelines.
\rangecap{11}~(11) use golden datasets or ground truth data for evaluation, an approach where models are tested against pre-verified, human-labeled data to calculate exact performance metrics: \participantQuote{We evaluate LLMs' response and score it out of 10, any score below 7 is considered a failure.}{P21}.
To track progress over time, participants use internal organizational benchmarking, assessing complex historical projects to determine whether newer models can successfully navigate difficult tool-calling workflows that caused previous iterations to fail.

\rangecap{14}~(14) engage in manual approaches like sanity checks, trial-and-error, and prompt testing to observe model behavior prior to full-scale integration.
The manual monitoring can also persist into production, as noted by P10: \participantQuote{There are occasional manual reviews from a sample size of \textins{production} results.}{P10}.
\rangecap{4}~(4) practitioners adopted an LLM-as-a-Judge approach, where a larger or more capable model acts as an evaluator to score the outputs of the primary integrated model against expected results.
 For deployments that participants characterized as security-critical, \range{3}~(3) practitioners perform red teaming and stress testing to identify untrusted inputs or malicious instructions hidden in code. Furthermore, in specialized industries, \rangecap{2}~(2) practitioners utilized User Acceptance Testing (UAT), where subject matter experts manually verify the correctness of the model's output to ensure it meets professional standards.
\subsubsection{Selection Actors} 
The actors involved in selecting and approving AI components vary widely across organizations, as does the need for formal approval prior to integration.

\myparagraph {Operational Size and Context}
\rangecap{8}~(8) participants indicated that in their larger organizations or regulated sectors (e.g., healthcare, fintech), those handling sensitive data need specific teams' approval.
As one participant noted: \participantQuote{The bigger company for which we were working, they had a separate three-hour meeting to analyze the entire flow, the entire architecture \textins{of a component}.} {P11}.
In such regulated settings, non-compliance with organizational or governmental policies can lead to rejection of a component.
Less regulated environments, such as startups or personal projects, typically impose fewer procedural constraints.
As another participant observed: \participantQuote{The government is more restricted \textelp{} but at the industry or startup level, they might not consider security that much}{P05}. In some cases, this governance manifests as informal guidance rather than enforceable rules: \participantQuote{We are advising developers to be careful with the cloud ones and avoid sharing very private things.}{P17}.

\myparagraph{Developer's Choice}
\rangecap{14}~(14) participants emphasized that they act as the primary, autonomous decision-makers for model selection: \participantQuote{It's usually me or the developer \textins{that} decide based on task, pricing, and other factors.}{P17}.
While highly regulated contexts enforce structured approval, non-regulated environments often delegate decision-making authority to project leaders or developers to select models based on their understanding of project requirements and risk considerations:\participantQuote{It's just myself who was involved \textelp{} I didn't have to consider more about the compliance.}{P09}.
Some teams also feel confident and equipped enough to know which model to choose, as mentioned by one participant: \participantQuote{Our engineers are equipped enough that they have worked on so many projects that they know the ups and downs, pros and cons of each model.}{P16}.
Overall, we observed that without formal regulatory oversight, model selection appears to be mostly determined and incorporated by practitioners' judgment rather than organizational policy.

\begin{tcolorbox}[colback=blue!5!white]
\textbf{Key Insight (Selection of Components):} The selection of AI as a software component is a multi-dimensional decision. Practitioners prioritize performance criteria such as cost, accuracy, logical reasoning, and use cases. However, these considerations are often constrained by infrastructure and hardware availability. Furthermore, the selection process is also governed by internal IT policies and driven by factors such as trust in the provider and their reputation, pre-existing institutional relationships or licensing, and adherence to approved workflows.
\end{tcolorbox}

\subsection{Challenges in AI Adoption (RQ2)}
In this section, we answer~\textit{RQ2: What challenges do practitioners face when integrating AI
Components?} Our questions targeted encountered challenges across four main areas: 
(1) technical hurdles and model limitations and 
(2) security challenges.
\subsubsection{Technical Hurdles and Model Limitations} 
Practitioners report a range of technical challenges and inherent model limitations that complicate the reliable integration of AI components into production systems.

\myparagraph{Model Performance}
\rangecap{15}~(15) participants noted that dealing with~\textit{AI slop} and the non-deterministic nature of AI models makes it difficult to achieve consistent, structured outputs required for production systems.
Participants reported that models produce biased, user-satisfactory hallucinations, deviate from the prompt, causing reliability issues: \participantQuote{We evaluate the risks and the type of output it gives. We need a model that can give consistent output.}{P12}.
\rangecap{12}~(12) participants consider performance and latency to be major pain points. Slow response times can negate the productivity benefits of using an LLM in the first place.
The constant API updates and volatility of AI component versions were mentioned by \range{7} (7) participants as an ongoing challenge, making it difficult to stabilize system behavior over time as stated by participant \participantQuote{It's always a bit challenging to find the right saddle point, you keep comparing models, and you don't know where to stop.}{P11}. Furthermore, \range{3}(3) practitioners pointed to the limitation regarding cross-linguistic model performance, noting that the models performed more accurately in English than in non-English contexts, alongside a lack of evaluation processes for non-English languages. As one participant mentioned, \participantQuote{we have a language barrier. Most of the evaluation pipelines and evaluation datasets are in English right now.}{P17}.

Beyond the initial selection challenge, practitioners also reported that provider-side updates introduce unpredictable behavioral changes that affect system reliability. \participantQuote{One challenge with language model providers is that models evolve quickly. Sometimes the provider upgrades the model, changes behavior slightly, or deprecates an old version. When that happens, the output can change.}{P19}. This version drift concern is discussed further as a security and supply chain challenge in Section 4.3.2.

\myparagraph{Integration with Existing Infrastructure}
\rangecap{10}~(10) of the participants reported systems needed to be custom-designed to interact with existing infrastructure and private ecosystems.
As a participant wished for their components: \participantQuote{Reducing the dependency of NVIDIA \textelp{} And something you can build into your own ecosystems.}{P03}.
Participants also mentioned that system design matters: \participantQuote{First, we focused heavily on the model capability itself, assuming better prompts or better models would solve most problems, but over time, we realized that the system design really matters.}{P19}.
Integrating models can rapidly generate technical debt: \rangecap{12} (12) of practitioners reported that orchestrating multiple tools, managing GPU memory usage, resolving dependency conflicts, and preventing local RAM crashes are major barriers.
\participantQuote{It can get complicated pretty quickly, creating a lot of technical debt.}{P07}.

\subsubsection{Security Challenges}
Practitioners identify several security-related challenges that arise from the integration of AI components, many of which stem from limited transparency, external dependencies, and a lack of established resources like documentation or guidance for secure integration or adoption of AI models.

\myparagraph{Vendor Lock-in} 
\rangecap{11}~(11) of participants expressed concerns regarding vendor lock-in, noting that reliance on a single LLM provider introduces systemic vulnerabilities. In particular, service disruptions, pricing shifts, or policy changes can significantly impact system stability and continuity. Related to vendor lock-in, practitioners also raised concerns about the security and reliability implications of unannounced model updates, which can lead to version drift. When a provider silently upgrades or deprecates a model version, system behavior can shift in ways that bypass existing safeguards or monitoring baselines, creating a supply chain risk distinct from traditional library versioning. \participantQuote{Sometimes the provider upgrades the model, changes behavior slightly, or deprecates an old version. When that happens, the output can change.}{P19}. This makes it difficult for practitioners to maintain consistent security assurances for models across the software development lifecycle.

\myparagraph{Blackbox Nature} 
\rangecap{7}~(7) participants reported limited transparency from providers regarding internal model behavior, described as a scenario where \textit{nobody’s talking about the middle}
which constrains their ability to fully assess risks associated with integrating the AI into their system. Consequently, their organizations default to evaluating models based solely on inputs and outputs, overlooking intermediate processes: \participantQuote{The approach to models is very sketchy. Not too many people are explaining what they're doing with models in their products.}{P07}.
This limited visibility into components complicates efforts of secure and trustworthy integration. 

\myparagraph{Lack of Documentation and Guidance} 
The majority of (14) participants highlighted a lack of formal guidance and insufficient documentation regarding the secure use and integration of AI components. This gap is notable because it affects practitioners at two distinct levels: the availability of model-level documentation from providers and the absence of organization-level guidance for secure integration practices. At the model level, incomplete or absent documentation forces practitioners to rely on behavioral observation and trial-and-error rather than informed security assessment~\participantQuote{Sometimes I don't choose a model because of a lack of documentation. I prefer a model that is well-documented, where you can see the code they use to train it, the benchmark, and the results.}{P02}. This shows that documentation gaps do not merely inconvenience practitioners, but they actively redirect selection decisions away from better-understood models, creating a systemic bias toward familiarity over rigor. At the organizational level, the absence of security-specific guidance leaves practitioners without actionable standards to follow, even when they recognize the need. \participantQuote{I didn't find much documentation regarding the security of LLMs, as most companies only began discussing these risks in detail within the last few months.}{P13}. Together, these two dimensions reveal that the lack of documentation is not simply a knowledge gap but a structural deficiency, a lack of institutional and provider-level resources. This manifests directly in calls for formal training, \participantQuote{I do believe there should be training for safe LLM usage, because in our case, our chatbot interacts with our services \textelp{} there should be training for safe LLM integration.}{P14}. Taken collectively, the findings indicate that guidance and documentation are not peripheral concerns but foundational prerequisites for secure AI integration that are currently unmet across the organizations.

\begin{tcolorbox}[colback=blue!5!white]
\textbf{Key Insight (Challenges):} Integrating AI components challenges traditional deterministic software practices.
The main difficulty lies not only in initial integration but in handling ongoing model volatility, silent API updates, and changing behavior.
Participants report accumulating technical debt, evaluation fatigue, and vendor lock-in as they work with opaque, non-deterministic models that are difficult to reliably assess and integrate into existing systems.
\end{tcolorbox}

\subsection{Perceived Risks and Applied Safeguards (RQ3)}
In this section, we present the findings for \textit{RQ3: How security risks are perceived and what safeguards are applied to mitigate the security risks?}
Our questions targeted perceived security risks and the safeguards implemented by practitioners.

\subsubsection{Perception of AI Security Risks}
Our participants' perceptions of AI security range from viewing vulnerabilities as unavoidable to highlighting the risks of human mismanagement.

\myparagraph{Data Exposure and Confidentiality Leakage}
\rangecap{22}~(22) participants indicated exfiltration of proprietary data or exposing Personally Identifiable Information (PII) as a perceived security concern. Participants expressed worry about sensitive data leaving secure corporate boundaries and reaching unintended parties. \participantQuote{\textins*{I}f sensitive data is involved, for example, PII or confidential documents, that will be a higher risk.}{P12}.
Similarly, another participant mentioned: \participantQuote{In a vendor compromise, an attacker hacks the model vendor, and the model becomes unreliable, as well as the data exposed.}{P19}. There is also a concern that model providers might use proprietary or client data to train their models and store it.
Concerns were also raised about data leaving the secure corporate environment or the vendor lock-in ecosystem and being accessed by third parties.
One participant cautioned: \participantQuote{There is a risk that you expose something that you shouldn't, and it goes into the wrong hands}{P21}.

\myparagraph{Inherent Vulnerability and Residual Risk}
\rangecap{11}~(11) practitioners viewed AI as intrinsically insecure due to its probabilistic nature.
Unlike traditional software libraries, which are viewed as deterministic behavior, selecting an AI model requires evaluating the probability of the outputs. As one of the participants stated, \participantQuote{LLMs are not plug-and-play and require careful design and monitoring.}{P12}, and another participant who characterized the experimental risks: \participantQuote{Security and AI have nothing to do with each other. AI is stochastic mathematics, while security, at its root level, is discrete mathematics.} {P05}
and also: \participantQuote{In my view, the risk is a residual risk. We can't mitigate it; it stays there forever.} {P07}. Other participants noted model hallucination risks in sensitive fields like AI mental health counseling, where random or incorrect answers could be dangerous.
Our participants were also concerned about models providing false positive results, prompt drifting, and eventual deprecation of models, which might require total replacement.

\myparagraph{Attack Vectors}
\rangecap{14}(14) participants mentioned prompt injection as a primary risk, where users can embed instructions in input documents to manipulate the model's logic or override agent interactions.
One participant shared an experience where a system broke after three days due to a flood of \textins{malicious} robo-requests originating from another continent.
Additionally, participants showed concern regarding vendor compromises, where an attacker might hack the model vendor itself, making the model unreliable and exposing all data.
\rangecap{4}(4) participants identified human mismanagement as a primary vulnerability: \participantQuote{If there's any specific security risk that concerns us, that would probably be the human factor. We somehow gave more access to the model than it actually needs, or authorization policies are misconfigured.}{P14}.
Finally, practitioners also showed concern towards a paradigm shift to ~\textit{Shadow AI}, and where AI is beginning to achieve autonomy, specifically regarding agents performing unauthorized operations in production environments or accessing critical codebases.

\subsubsection{Adopted Safeguards}
Our participants employ a range of safeguards to mitigate risks associated with AI integration, though their application is often ad hoc and shaped by perceived risk and organizational constraints.

\myparagraph{Data Handling}
\rangecap{22}(22) participants stated that PII filtering is a key safeguard for protecting sensitive data in their AI systems.
They use preprocessing pipelines (e.g., rule-based patterns or NER models) to detect and redact identifiers such as names, emails, and IDs before sending data to external models:
\participantQuote{\textins{We use} a dedicated layer that checks both prompts and responses for PII before and after model interaction.}{P02}. 
Practitioners also use data masking and synthetic data generation to replace sensitive values, particularly in highly regulated sectors like healthcare.
Additionally, participants take it upon themselves to anonymize datasets locally before exporting them to any AI tools.
\rangecap{8}(8) participants preferred deploying open-source models internally to ensure data stays in-house and does not expose information to outside cloud environments.
Our participants mention ecosystems such as Azure Foundry, AWS Bedrock, or Google Vertex AI, which provide a gateway to models while ensuring data remains within the company's specific cloud region.
Additionally, practitioners pointed out that models and chat sessions are hosted in isolated environments with no external internet access to prevent data exploitation. 

\myparagraph{Monitoring and Reliability}
\rangecap{14}(14) practitioners mentioned monitoring tools (e.g., LangSmith, LangFuse) for operational observability, including token usage tracking and prompt versioning.

Participants are also using ~\textit{LLMs-as-judges} to assess the primary model’s outputs:\participantQuote{Another LLM, which basically acts as a judge for the responses, or the differences between the actual output and the expected output.}{P16}.
For \range{18}(18) participants, perceptions of security are heavily influenced by the trustworthiness of model providers: \participantQuote{I try my best to look for the ones that are one directly from notable companies and use that.} {P10} and \participantQuote{Hugging Face is kind of trust\textins*{ed}. It's a trusted place every developer goes to get a model's weight.}{P02}.
Trust is also related to how the vendors handle and potentially use customer data for training: \participantQuote{If you're going for paid ones, the main security thing is, talk to the provider, make sure your data is not used for any training purpose.}{P04}.
Some participants rely on their organizational IT team's approval: \participantQuote{If it is an IT-approved model, we are good to use it, and we don't need to go ahead and verify it ourselves.} {P21}.
Overall, similar to classic dependencies, the trust in the provider or community appears to be used as a general proxy for model security. 

\begin{tcolorbox}[colback=blue!5!white]
\textbf{Key Insight: Risks and Safeguards} Practitioners' major perceived risks are data exposure and confidentiality. To mitigate risks, practitioners adopted several data protection measures to prevent the leakage of sensitive data. Additionally, model adopters heavily trust and rely on model vendors as a security stand-in. 
\end{tcolorbox}

%% file: Tables_and_Figures/Scale_diagram.tex
\begin{figure}[tbp]
    \centering
    % Resizes the entire TikZ picture to exactly match the column width
    \resizebox{\columnwidth}{!}{%
    \begin{tikzpicture}[scale=1.5, transform shape, x=0.75cm, y=1cm, font=\Large]
        
        % Main horizontal line (15cm wide for drawing scale)
        \draw (0,0) -- (15,0);

        % Solid vertical end lines
        \draw (0,-0.8) -- (0,0.8);
        \draw (15,-0.8) -- (15,0.8);

        % Dotted vertical dividers mapped to percentage values
        % 15% -> 2.25, 30% -> 4.5, 45% -> 6.75, 55% -> 8.25, 70% -> 10.5, 85% -> 12.75
        \foreach \x in {2.25, 4.5, 6.75, 8.25, 10.5, 12.75} {
            \draw[densely dotted] (\x,-0.8) -- (\x,0.8);
        }

        % Top labels (Numerical Ranges scaled for 22 participants)
        \node at (-0.7, 0.45) {0};
        \node at (1.125, 0.45) {1--3};
        \node at (3.375, 0.45) {4--6};
        \node at (5.625, 0.45) {7--9};
        \node at (7.5, 0.45) {10--12};
        \node at (9.375, 0.45) {13--15};
        \node at (11.625, 0.45) {16--18};
        \node at (13.875, 0.45) {19--21};
        \node at (15.7, 0.45) {22};

        % Middle labels (Qualitative Terms)
        \node at (-0.7, -0.45) {None};
        \node at (1.125, -0.45) {A few};
        \node at (3.375, -0.45) {Some};
        \node at (5.625, -0.45) {Many};
        \node[align=center] at (7.5, -0.45) {About\\[-0.5ex]Half};
        \node at (9.375, -0.45) {Majority};
        \node at (11.625, -0.45) {Most};
        \node[align=center] at (13.875, -0.45) {Almost\\[-0.5ex]All};
        \node at (15.7, -0.45) {All};

        % Bottom labels (Percentages)
        \node[anchor=north] at (0, -0.8) {0\%};
        \node[anchor=north] at (2.25, -0.8) {15\%};
        \node[anchor=north] at (4.5, -0.8) {30\%};
        \node[anchor=north] at (6.75, -0.8) {45\%};
        \node[anchor=north] at (8.25, -0.8) {55\%};
        \node[anchor=north] at (10.5, -0.8) {70\%};
        \node[anchor=north] at (12.75, -0.8) {85\%};
        \node[anchor=north] at (15, -0.8) {100\%};

    \end{tikzpicture}%
    }
     \vspace{-15pt}
    \caption{Terminology used to report the ranges of results.}
    \label{fig:interview_ranges}
\end{figure}

%% file: Tables_and_Figures/Participant_table.tex
\begin{table*}[htbp]
\centering
\caption{Participant Demographics}
\label{tab:demogarphics}
\begin{threeparttable}% \begin{tabular}{|l|l|l|l|l|l|}

\begin{tabular}{@{}
c
>{\centering\arraybackslash}p{3.5cm}
c
>{\centering\arraybackslash}p{3.5cm}
>{\centering\arraybackslash}p{2cm}
>{\centering\arraybackslash}p{2cm}
c
@{}}
\toprule

\textbf{Participant ID} & \textbf{Role} & \textbf{Years of Experience} & \textbf{Organization Type\tnote{†}}    & \textbf{Codes} & \textbf{Duration} \\ \midrule
P01            & Senior Software Developer & 4-6 years           & Cybersecurity        & 64    & 1:34:43\tnote{§}               \\ 
P02            & Machine Learning Engineer & 7-10 years          & Finance              & 52    & 0:45:05               \\ 
P03            & Senior AI Engineer        & Over 10  years      & Manufacture          & 50    & 0:59:19               \\ 
P04            & Software Engineer         & 4-6 years           & Cybersecurity        & 48    & 0:56:07               \\ 
P05            & AI Engineer               & 7-10 years          & Finance              & 55   & 0:45:44               \\ 
P06            & AI Engineer               & 4-6 years           & Finance              & 78   & 0:59:08               \\ 
P07            & CEO                       & Over 10 years       & Cybersecurity        & 60    & 1:25:36\tnote{§}               \\ 
P08            & Project Manager           & 4-6 years           & Education            & 57    & 0:49:34               \\ 
P09            & Software Developer        & 1-3 years           & Education            & 86    & 1:00:06\tnote{§}               \\ 
P10            & AI Engineer             & 4-6 years              & Finance             & 69    & 1:00:22               \\ 
P11            & Software Developer        & 1-3 years           & Finance              & 70    & 1:05:28\tnote{§}               \\ 
P12            & Software Developer        & 4-6 years           & Finance              & 79    & 0:45:52               \\ 
P13            & Software Engineer         & 4-6  years          & Cybersecurity        & 92    & 0:48:41               \\ 
P14            & Software Engineer         & 1-3 years           & Cybersecurity        & 71    & 1:00:02               \\ 
P15            & Software Developer        & 1-3 years           & Healthcare           & 90   & 0:46:04               \\ 
P16            & AI Engineer               & 4-6 years           & Healthcare           & 80   & 1:20:30\tnote{§}               \\ 
P17            & Senior Data Scientist     & 4-6 years           & Finance              & 71    & 0:46:42               \\ 
P18            & Data Scientist            & 1-3 years           & Education            & 79    & 0:48:54               \\ 
P19            & AI Architect              & Over 10 years       & Manufacturing        & 105    & 1:04:19\tnote{§}               \\ 
P20            & Senior Data Scientist     & 4-6 years           & Manufacturing        & 74    & 1:18:49\tnote{§}               \\ 
P21            & Software Engineer         & 1-3 years           & Cybersecurity        & 98    & 1:04:55\tnote{§}               \\ 
P22            & CEO                       & Over 10 years       & Software Development & 59   & 0:41:48               \\ 
\bottomrule
\end{tabular}

\begin{tablenotes}
\footnotesize
\item[†] Binned self-reporting of participants to preserve their privacy.
\item[§] Participant voluntarily exceeded planned interview time.
\end{tablenotes}
\end{threeparttable}
\end{table*}

%% file: sections/06-discussion.tex
In the following section, we synthesize our findings into key themes and discuss their implications for the secure integration of AI components within the software supply chain. 
We then discuss recommendations for AI adopters and practitioners, AI providers, and researchers to improve the security and governance of AI integration in software development.

\subsection{Identified Themes or Back to the Future}

Based on our findings, we argue that, with AI models now embedded as critical components of software systems, the industry is facing a \textit{Back to the Future} moment in the software supply chain.
Current AI adoption practices mirror earlier software dependencies, where rapid reuse and readily available components were favored while their role in the software supply chain and associated attack surfaces were overlooked~\cite{tanzil2024people, lysenko2025select}, an issue that took years of research and practice to address and remains ongoing. 
In the following, we discuss how these dynamics reappear in the integration of AI models and their implications for security.

\myparagraph{Rapid Adoption and Results over Security}
Several participants of our study emphasized rapid and exploratory adoption of new AI technologies over security considerations:
\participantQuote{I tend not to focus too much on users' or enterprises' security concerns, instead aggressively adopt new technologies, seeing where it takes me.}{P05}. 
Security is often treated as an afterthought or actively de-prioritized in favor of performance, cost-efficiency, and rapid market release~\cite{murat2024overview,klostermeyer2024skipping}. Similarly, our participant mentioned focusing \participantQuote{less on the security, more on the results.}{P22}.
Despite awareness, many of our participants still consider security a secondary or tertiary concern: \participantQuote{Security is less of a priority.}{P2}. 

This is not simply a matter of individual oversight, but rather reflects a structural dynamic in which time-to-market pressure, cost sensitivity, and competitive positioning consistently outweigh security investment at the decision-making level. This pattern closely mirrors what prior work has documented in traditional software dependency management,  where developers prioritized reuse and availability over security scrutiny for decades~\cite{gkortzis2021software, gkortzis2021software} before the consequences became unavoidable. Echoing the traits the participant mentioned,
\participantQuote{In my opinion, stop focusing on all the security and let's focus on what the thing can do, right? What can this project do? What is the capability of the system?}{P22}.This practice attitude also extends to regional and cultural dimensions. Participants associating security consciousness related to geographic location influence stating that \participantQuote{\textins*{S}ecurity is a very America-focused issue.}{P05} continuing \participantQuote{In China, they'd likely discuss building AI girlfriends. Only in American.. especially from East side people wonder how to combine AI and security.}{P05}. These findings present a need for security to be treated as a design requirement in AI integration guidance, rather than an optional concern left to individual or organizational discretion.

\myparagraph{Trust as a Proxy for Security in Model Selection}
Similar to the selection of open-source software packages~\cite{wermke2023always}, our participants rely on coarse-grained signals of trust when selecting AI models, such as popularity metrics, community activity, or affiliation with major technology providers. This behavior is structurally analogous to how developers historically selected open-source packages based on download counts and GitHub stars~\cite{reichert2024software}, rather than security audits. \participantQuote{[I] don't have that time for a formal or organizational level to go into each and every architecture and everything. It all depends on Stack Overflow upvotes; it depends on community things.}{P11}.

Further, we observed that trusting major AI and cloud providers is the most common approach and assumed secure choice for most organizations, as participants noted,\participantQuote{A lot of trust goes on the cloud vendors.}{P11}. However, 
studies have shown that trust alone is insufficient for security in software engineering due to unclear trust concepts and potential security vulnerabilities~\cite{khati2025mapping}.
AI-generated output is susceptible to vulnerabilities, posing challenges for secure software development \cite{vieira2025leveraging,10795572,ramirez2024state}.
% Often leads to trade-offs, where a technically superior model might be overlooked in favor of one that offers better security guarantees or fits within an established compliance perimeter.
While this offloads the burden of data residency, it limits architectural flexibility and concentrates systemic cyber risk among a few tech entities, mirroring broader concerns regarding software supply chain vulnerabilities in AI~\cite{OWASP03,colares2026multi}.
This highlights the importance of an approach where the model incorporates a comprehensive evaluation that integrates both performance and security
aspects, along with trust, to guide stakeholders in choosing the most suitable
models according to their needs.

\myparagraph{Not One Size Fits All}
We found that security considerations in AI model integration as components are not uniform and vary significantly across contexts.
The observed differences in security posture are shaped by participants’ roles, organizational maturity, and regulatory environment.
Practitioners in senior roles or highly regulated sectors such as finance, healthcare, and supply chain management mentioned greater security concerns.
In these settings, requirements such as data residency, compliance with regulations like GDPR and HIPAA, and enterprise-grade security guarantees are treated as non-negotiable in model selection.
On the other side, for many developers, particularly those in startups or less-regulated sectors, security was described as a \textit{lower priority} or a concern handled by other teams or at a later stage: \participantQuote{The government is more restricted \textelp{} But for private companies so far. Their data isn't as sensitive, and they don't care.}{P06}.
Their immediate drivers for LLM selection and integration were mainly cost: \participantQuote{The only guideline might be the costing.}{P06}.
Further, the lack of security considerations is not absolute but is instead a cultural and structural issue. Additionally, the participant reinforced the statement by mentioning that~\participantQuote{Leading government projects might require more security measures and training on how to use AI or something. But at the industry or startup level, they might not consider this much.}{P05}.
For some organizations, decisions are routed through dedicated IT, legal, and compliance teams, and model selection is heavily influenced by partnership or licensing offered by platforms.
However, in organizations lacking a strong DevSecOps culture, security responsibility is often delegated, leaving developers to focus on implementation while assuming security is not their responsibility.
This aligns with existing research on software development practices, which indicates that security is often neglected unless it is explicitly integrated into the development lifecycle and supported by organizational mandates~\cite{knapp2009security, raghavan2017integrative}.

Moreover, the \textit{move fast and break things} ethos, while common in software development, is particularly perilous in the context of AI components, where the attack surface is novel and not fully understood. 
We also observed regional variations in security prioritization.
Factors such as regulatory frameworks, market pressures, and divergent philosophical approaches to technology adoption shape these strategies.
In markets prioritizing rapid innovation, security concerns may be deprioritized, whereas in highly regulated environments, security is often treated as a foundational requirement. This suggests that external mandates, rather than intrinsic motivation, are currently the primary driver of secure AI adoption. These findings show that security is consistently integrated into development practices only when it is institutionally mandated.

\myparagraph{AI Security as a Software Supply Chain Problem}

The traditional software supply chain is built on dependencies with relatively deterministic behavior and vulnerabilities that are often cataloged in databases (e.g., CVEs)~\cite{CVE}, enabling structured risk assessment and patch management. But integrating AI components introduces a fundamentally different, dynamic upstream component. We found that practitioners perceive AI selection not as a simple choice of a library, but as a strategic sourcing decision, akin to engaging a critical new vendor: \participantQuote{When you're choosing a package or a library, the performance is very deterministic. But with LLMs, it's very difficult because there are so many use cases, and the results, even from the actual model providers, vary a lot when you're actually testing them in the real world.}{P16}.
Practitioners mentioned turning to open-source models to manage costs and avoid vendor lock-in.
However, this path has trade-offs, which require practitioners to maintain hardware constraints, secure on-premise infrastructure, and verify model origin.
Our findings echo the idea that AI security cannot be achieved solely through pre-deployment alignment or static benchmarks~\cite{cao2025survey,verma2024operationalizing}. Practitioners emphasized challenges such as provider rate limiting, model version drift, policy changes, and deprecations. These issues map directly to the AI model lifecycle and supply-chain risks identified in security literature~\cite{williams2025research, wang2025large, wu2402new,hu2025large}.
The focus on operational resilience extends the existing view of AI supply-chain security, which often considers threats such as data poisoning, jailbreaking, and backdoors.
Practitioners were more concerned with business continuity, needing to plan for model deprecations, maintain abstraction layers to swap providers, and design observability pipelines that are resilient to changing APIs or safety policies.
At present, the AI ecosystem supply chain infrastructure can be considered largely ad hoc and lacking metrics such as risk scores, vulnerability databases, and SBOM-like artifacts.

\myparagraph{Uncontrolled AI Integration}
The governance challenge is further exacerbated by the rise of \textit{Shadow AI}, where developers use artificial intelligence tools, applications, or models within an organization without authorization or oversight from IT and security teams. Our participants described that most of the time, developers independently adopt AI models and integrate them into production systems without formal approval processes, particularly in less-regulated environments.
The trend mirrors the historical challenges of \textit{Shadow IT}, reflecting a broader reality where the democratization of AI tools consistently outpaces the development of security policies.
Failing to control model interactions leads to significant \textit{AI technical debt}~\cite{sculley2015hidden, valence, capture}. Shadow IT was ultimately addressed not by prohibition but by developing governance frameworks that made sanctioned alternatives accessible~\cite{gyory2012exploring,shadow-it}. A similar approach is needed for AI organizations to invest in approved AI integration architectures and internal governance policies that reduce the friction of secure adoption, rather than relying on developer self-restraint alone.

\myparagraph{AI Threats and Defensive Gaps}
Our participants identified a range of AI-specific threats, including data leakage, prompt injection, and unintended model behavior, and demonstrated baseline awareness of these risks.
In practice, however, defenses largely rely on traditional measures such as input sanitization, PII filtering, and perimeter controls, alongside emerging but still immature approaches like LLM-as-a-Judge and ad hoc monitoring.
We consider this a gap where recognized risks are addressed with controls that are not well-suited to the non-deterministic and compositional nature of AI components.
 
For instance, the observation that \participantQuote{you should never build your entire system fully on AI.}{P01} reflects a cautious approach that treats AI as an augmentation rather than a fully trusted component.
Many organizations continue to rely on traditional perimeter-based security measures, such as basic input sanitation or PII stripping.
These methods are, however, insufficient against novel attack vectors such as indirect prompt injection and adversarial jailbreaking~\cite{greshake2023not}.
AI models cannot be reliably protected by simple input filtering, as the model inherently conflates system instructions with untrusted user data~\cite{perez2022ignore}.
The reliance on legacy security paradigms for non-deterministic, next-generation agents constitutes a critical vulnerability in current industry practice~\cite{greshake2023not}.
Concurrently, the emergence of the \textit{LLM-as-a-judge}, which leverages AI to evaluate other AI outputs, represents a shift toward adaptive, model-driven evaluation. The process offers scalability and flexibility with improved performance. Nevertheless, this approach introduces several barriers, including sensitivity to prompt design, biases, and ethical concerns~\cite{szymanski2025limitations}. 
Studies advocate for hybrid evaluation frameworks that integrate AI models' judgments with human oversight or complementary metrics to ensure validity and reliability~\cite{chen2024humans,bavaresco2025llms,gu2024survey}.

\subsection{Recommendations}
Based on our findings, we outline recommendations for future work for AI adopters and practitioners, AI providers, and researchers to address observed gaps in security practices, improve model selection and integration processes, and strengthen the secure AI supply chain.

\subsubsection{AI Adopters/ Practitioners}
The insights from our participants underscore that integrating LLMs securely requires a fundamental shift from a reactive to a proactive security posture.
Practitioners need to move beyond viewing the model as a simple API call and instead treat it as a critical component with its own unique threat surface and supply chain implications.

\myparagraph{Treat AI Model Selection as a Supply Chain Decision}
When integrating AI components, practitioners need to consider the model as a critical link in the software supply chain.
Practitioners should consider the model's transparency and provenance, including its training data origins and data-handling processes.
Additionally, a compliance assessment may be required to align with relevant legal, regulatory, data retention, and organizational governance requirements.
Model providers' security practices also play a role, including the frequency and reliability of security updates and patches, the availability of vulnerability disclosure, and the availability of community or experts' security scrutiny.
Together, these factors would help enable a security-informed approach to model integration.

\myparagraph{Prioritize Security-by-Design} Security should be treated as a core design requirement when integrating AI components, rather than solely a post-integration concern. Improving data sovereignty is necessary for data control and compliance with regulatory governance.

\myparagraph{Enhance Privacy and Policies} 
Before model integration, strong privacy protections, including access control, encryption, and provenance tracking, should be established.
Organizations should also formalize model integration policies by embedding security requirements across the lifecycle, including pre-adoption risk assessment, data governance controls, and monitoring with a human in the loop.

\myparagraph{Implement Layered Defenses}
Practitioners should adopt a defense-in-depth strategy rather than relying on a single safeguard.
These could be combining mechanisms such as structured prompting, architectural separation, least-privilege access, and policy checks for high-impact operations. 
Additionally, models should be adopted based on their compatibility with downstream safeguards, such as input/output filtering, guardrails, and monitoring tools, to enable cohesive, secure system integration.

\subsubsection{AI Provider}

Providers and developers of AI models are uniquely positioned to raise the industry-wide security baseline.
The challenges and workarounds described by participants show a need for platforms that embed security as a native, transparent, and standardized feature, rather than leaving the implementation burden entirely on the adopter.

\myparagraph{Enhance Transparency and Auditability}
Providers should make model lineage and behavior transparent.
Offering detailed logs and audit trails allows adopters to investigate data flows.
Similar to a Software Bill of Materials (SBOM), providers could include an ~\textit{AI Bill of Materials} that details training data and other model components.

\myparagraph{Standardize APIs and Schemas}
Consistency reduces configuration errors.
By standardizing API interfaces and output formats, providers can make it easier for adopters to implement robust validation and parsing, thereby reducing the surface area for exploits.
  
\myparagraph{Strengthen Built-in Security Features:} 
Participants are currently building their own safeguards and custom monitoring engines.
This duplicates effort and leads to inconsistent security levels across the industry.
Providers should offer integrated safeguard or monitoring facilities capabilities as native features of their SDKs or APIs.
Included tools that automatically scan prompts for credentials before transmission or block PII by default, allowing adopters to secure their applications with minimal overhead.
  
\myparagraph{Address Data Sovereignty Concerns}
AI model providers should protect user privacy and provide zero-retention data policies, or the ability to opt out of model training or geographically restricted data processing.
Such authority will help in ensuring that the model does not inadvertently become a repository for sensitive information.

\subsubsection{Researchers}
The experiences of practitioners highlight several areas where academic research can provide the foundational knowledge and tools needed to secure the next generation of
AI components.
The current landscape is characterized by ad-hoc evaluation methods and a poor understanding of emergent risks, presenting opportunities for structured, empirical investigation.

\myparagraph{Standardize LLM Security Frameworks and Benchmarks}
The industry lacks a common language for measuring security and safety evidence.
As such, most participants rely on manual tests and an internal dataset.
Researchers could establish universal, reproducible benchmarks that allow developers and adopters to quantitatively compare the security posture of different models.
 
\myparagraph{Investigate Black Box Vulnerabilities and Explainable AI Security}
Since many models are proprietary black boxes, research is needed into methods that infer internal vulnerabilities from external behaviors.
Advancing explainable AI for security-critical decisions would align with the practical need for explainability and auditability, which one participant described as a \enquote{ hard blocker} for deployment in regulated environments.

\myparagraph{Impact of AI Autonomy on Cybersecurity}
Research needs to move beyond static model analysis to understand how increased agency changes the threat landscape.
Practitioners are moving beyond simple chatbots to build complex, agentic systems and are aware of the need to allow them to perform autonomous operations without human oversight. Research could investigate new paradigms for autonomous agents sp that they are less susceptible for malicious purposes and how their interactions within a complex ecosystem of plugins and enterprise software can be improved.

\myparagraph{Human-centered Studies of AI Security and Policy}
There is a need for human-centered and longitudinal studies of AI security work.
Research could investigate how practitioners actually interact with AI security controls to understand why safeguards are adopted or ignored.
Longitudinal studies are also useful to track how security risks evolve as models are updated, fine-tuned, and chained together over time.
Such human-centered approaches will be necessary for developing security solutions that are not only technically robust but also practically effective.

%% file: sections/07-conclusion.tex
The increasing integration of AI components into modern software has fundamentally transformed the software development landscape, with models becoming a foundational dependency in the software supply chain.
In this semi-structured interview study with 22 software practitioners, we explored the current operational strategies and security posture of their AI integrations.

 Our findings reveal that the industry is repeating historic software supply chain mistakes (\textit{Back to the Future}) by prioritizing functionality, cost, and rapid deployment over security and component choices.
 Security considerations in integrating AI into software development are often treated as secondary or tertiary requirements.
 Because developers treat AI as black boxes, the security burden has been entirely displaced onto reactive infrastructure-level safeguards and trust in the vendor, leaving systems vulnerable to compounding technical debt and vendor lock-in.
 As AI components evolve from passive text generators to autonomous agents capable of executing system-level actions, perimeter-based defenses like input filtering will no longer suffice.
 The industry must transition to a proactive, security-by-design approach that elevates security to a fundamental requirement alongside performance.